\documentclass[preprint,showpacs,preprintnumbers,amsmath,amssymb]{revtex4}

\usepackage{graphicx}
\usepackage{dcolumn}
\usepackage{bm}

\begin{document}

\preprint{APS/123-QED}

\title{ Isospin effects on fragment production\\}

\author{Anupriya Jain}
\author{Suneel Kumar}%
 \email{suneel.kumar@thapar.edu}
\affiliation{%
School of Physics and Material Science, Thapar University, Patiala-147004, Punjab (India)\\
}%

\date{\today}

\begin{abstract}
To understand the isospin effects in nearly symmetric nuclear matter we performed a complete systematical theoretical study within an Isospin dependent Quantum Molecular Dynamical Model (IQMD) and using Minimum Spanning Tree (MST) algorithm. Simulations are carried out for the reactions $^{124}Sn_{50}+ ^{124}Sn_{50}$ and $^{107}Sn_{50}+ ^{124}Sn_{50}$. The collision geometry is varied from central to peripheral. We find that neutron rich colliding nuclei are better candidate to study the isospin effects.\\
\end{abstract}

\maketitle
\baselineskip=1.5\baselineskip\
\section{Introduction}
Heavy-ion collisions have been extensively studied over the last decades.
The behavior of nuclear matter under the extreme conditions of temperature, density, angular momentum etc., is a very important aspect of heavy-ion physics. One of the important quantity which has been used extensively to study this hot and dense nuclear matter is the collective 
transverse in-plane flow \cite{1}. This quantity vanishes at a certain incident energy. Finite nuclei studies predict values for the symmetry energy at saturation of the order of 30-35 MeV. In heavy ion collisions highly compressed matter can be formed for short time scales, thus the study of such a dynamical process can provide useful information on the high density symmetry energy. Even at low incident energies which belong to even smaller baryonic densities, the isospin dependence of the mean field potential was shown to yield same result obtained with potentials that has no isospin dependences. These results are in similar lines and it also indicates that even binary phenomena like fission will also be insensitive towards isospin dependence of the dynamics \cite{2}. Recently theoretical studies on the high density symmetry energy have been started by investigating heavy ion collisions of asymmetric systems \cite{3,4}. Comparisons of collisions of neutron-rich to that of neutron-deficient systems provide a means of probing the asymmetry term experimentally \cite{5,6,7}. The experimental analysis of the isospin effects on fragment production has yielded several interesting observations: Dempsey et al. \cite{8} in their investigation of 
$^{124,136}Xe+^{112,124}Sn $ at 55MeV/ nucleon found that multiplicity of IMF's increases with the neutron excess of the system. A more comprehensive study was carried out by Buyukcizmeci et al. \cite{9} showed that symmetry energy  of the hot fragments produced in the statistical freeze-out is very important for isotope distributions, but its influence is not very large on the mean fragment mass distributions.  Symmetry energy effect on isotope distributions can survive after secondary de-excitation. Moreover Schmidt et al. \cite{10} in their investigation on the analysis of LCP's production and isospin dependences of $^{124}Sn+^{64}Ni$, $^{124}Sn+^{58}Ni$, $^{124}Sn+^{27}Al$ at 35MeV/nucleon and 25MeV/nucleon collisions found that isospin effects were demonstrated in the observables, such as the angular distribution of light particles emitted in central collisions at 35MeV/nucleon and LCP's emission. On the other hand Tsang et al. 
\cite{11} in their investigation of $^{112}Sn + ^{124}Sn$, $^{124}Sn + ^{112}$Sn systems at an incident energy of E=50 MeV/nucleon showed the effects of isospin diffusion by investigating heavy-ion collisions with comparable diffusion and collision time scales. They showed that the isospin diffusion reflects driving forces arising from the asymmetry term of the EOS. With the passage of time, isospin degree of freedom in terms of symmetry energy and nucleon-nucleon cross section is found to affect the balance energy or energy of vanishing flow and related phenomenon in heavy-ion collisions \cite{12}. Our present study will shed light on isospin effects on multiplicity of fragments. We present microscopic predictions of nucleon-nucleon cross section in isospin asymmetric nuclear matter. In asymmetric matter, the cross section becomes isospin dependent. It depends on the relative proton and neutron concentrations, which of course also implies that the pp/nn and the np cases will in general be different from each other. \\
This study is done within the framework of isospin-dependent quantum molecular dynamics model
that is explained in section-II. The results are presented in section-III. We present summary in section-IV.\\

\section{ISOSPIN-DEPENDENT QUANTUM MOLECULAR DYNAMICS (IQMD) MODEL}

Theoretically many models have been developed to study the heavy ion collisions at intermediate energies. One of them is quantum molecular dynamical model (QMD) \cite{13,14}, which incorporates N-body correlations as well as nuclear EOS along with important quantum features like Pauli- blocking and particle production.\\
In past decade, several refinements and improvements were made over the original QMD. The IQMD \cite{15} model overcomes the difficulty as it not only describe the ground state properties of individual nuclei at initial time but also their time evolution. In order to explain experimental results in much better way and to describe the isospin effect appropriately, the original version of QMD model was improved which is known as isospin-dependent quantum molecular dynamics (IQMD) model.\\
The isospin-dependent quantum molecular dynamics (IQMD)\cite{15} model treats different charge states of
nucleons, deltas and pions explicitly, as inherited from the VUU model. The IQMD model has been used successfully
for the analysis of large number of observables from low to relativistic energies. The isospin degree of
freedom enters into the calculations via both cross-sections and mean field.\\
In this model,baryons are represented by Gaussian-shaped density distributions
\begin{equation}
f_i(\vec{r},\vec{p},t) = \frac{1}{\pi^2\hbar^2}~e^{-(\vec{r}-\vec{r_i}(t))^{2}\frac{1}{2L}}~e^{-(\vec{p}-\vec{p_i}(t))^{2}\frac{2L}{\hbar^2}}.
\end{equation}
Nucleons are initialized in a sphere with radius $R= 1.12 A^{1/3}$ fm, in accordance with the liquid drop model. Each nucleon occupies a volume of $h^3$, so that phase space is uniformly filled. The initial momenta are randomly chosen between 0 and Fermi momentum($p_F$). The nucleons of target and projectile interact via two and three-body Skyrme forces and Yukawa potential. The isospin degree of freedom is treated explicitly by employing a symmetry potential and explicit Coulomb forces
between protons of colliding target and projectile. This helps in achieving correct distribution of protons and neutrons
within nucleus.\\
The hadrons propagate using Hamilton equations of motion:
\begin{equation}
\frac{d{r_i}}{dt}~=~\frac{d\it{\langle~H~\rangle}}{d{p_i}}~~;~~\frac{d{p_i}}{dt}~=~-\frac{d\it{\langle~H~\rangle}}{d{r_i}},
\end{equation}
with
\begin{eqnarray}
\langle~H~\rangle&=&\langle~T~\rangle+\langle~V~\rangle\nonumber\\
&=&\sum_{i}\frac{p_i^2}{2m_i}+
\sum_i \sum_{j > i}\int f_{i}(\vec{r},\vec{p},t)V^{\it ij}({\vec{r}^\prime,\vec{r}})\nonumber\\
& &\times f_j(\vec{r}^\prime,\vec{p}^\prime,t)d\vec{r}d\vec{r}^\prime d\vec{p}d\vec{p}^\prime .
\end{eqnarray}
 The baryon-baryon potential $V^{ij}$, in the above relation, reads as:
\begin{eqnarray}
V^{ij}(\vec{r}^\prime -\vec{r})&=&V^{ij}_{Skyrme}+V^{ij}_{Yukawa}+V^{ij}_{Coul}+V^{ij}_{sym}\nonumber\\
&=& \left [t_{1}\delta(\vec{r}^\prime -\vec{r})+t_{2}\delta(\vec{r}^\prime -\vec{r})\rho^{\gamma-1}
\left(\frac{\vec{r}^\prime +\vec{r}}{2}\right) \right]\nonumber\\
& & +~t_{3}\frac{exp(|\vec{r}^\prime-\vec{r}|/\mu)}{(|\vec{r}^\prime-\vec{r}|/\mu)}~+~\frac{Z_{i}Z_{j}e^{2}}{|\vec{r}^\prime -\vec{r}|}\nonumber\\
& & + t_{6}\frac{1}{\varrho_0}T_3^{i}T_3^{j}\delta(\vec{r_i}^\prime -\vec{r_j}).
\label{s1}
\end{eqnarray}
Here $Z_i$ and $Z_j$ denote the charges of $i^{th}$ and $j^{th}$ baryon, and $T_3^i$, $T_3^j$ are their respective $T_3$
components (i.e. 1/2 for protons and -1/2 for neutrons). Meson potential consists of Coulomb interaction only.
The parameters $\mu$ and $t_1,.....,t_6$ are adjusted to the real part of the nucleonic optical potential. For the density
dependence of nucleon optical potential, standard Skyrme-type parameterization is employed.
The choice of equation of state (or compressibility) is still controversial one. Many studies
advocate softer matter, whereas, much more believe the matter to be harder in nature. 
We shall use both hard (H) and soft (S) equations of state that have 
compressibilities of 380 and 200 MeV, 
respectively.\\

The binary nucleon-nucleon collisions are included by employing the collision 
term of well known VUU-BUU equation. The binary collisions
are done stochastically, in a similar way as are done in all transport models. During the propagation, two nucleons are
supposed to suffer a binary collision if the distance between their centroids
\begin{equation}
|r_i-r_j| \le \sqrt{\frac{\sigma_{tot}}{\pi}}, \sigma_{tot} = \sigma(\sqrt{s}, type),
\end{equation}
"type" denotes the ingoing collision partners (N-N, N-$\Delta$, N-$\pi$,..). In addition,
Pauli blocking (of the final
state) of baryons is taken into account by checking the phase space densities in the final states.
The final phase space fractions $P_1$ and $P_2$ which are already occupied by other nucleons are determined for each
of the scattering baryons. The collision is then blocked with probability
\begin{equation}
P_{block}~=~1-(1-P_1)(1-P_2).
\end{equation}
The delta decays are checked in an analogous fashion with respect to the phase space of the resulting nucleons.\\

\section{Results and Discussion}
\begin{figure}
\includegraphics{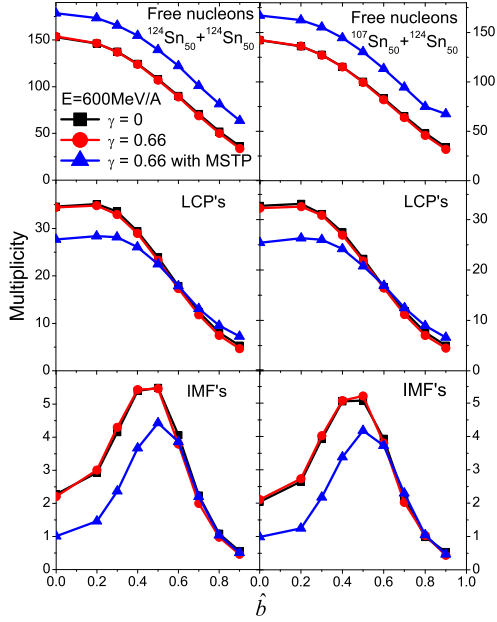}
\caption{\label{Fig:1} Multiplicity of free nucleons, LCP's and IMF's as a function of scaled impact parameter.}
\end{figure}

For the present study we have simulated $_{50}Sn^{124}+_{50}Sn^{124}$  and $_{50}Sn^{107}+_{50}Sn^{124}$   reactions by using isospin dependent quantum molecular dynamics (IQMD) model  at incident energy 600 MeV/nucleon at scaled impact parameters ( 0.0, 0.2, 0.3, 0.4, 0.5, 0.6, 0.7, 0.8, 0.9). The collision geometry used is from central to peripheral one. The phase space generated using IQMD model is analyzed by using minimum spanning tree [MST] algorithm \cite{16} and minimum spanning tree with momentum cut \cite{16} [MSTP]. The results obtained are discussed as: \\
 In Fig.\ref{Fig:1},shows multiplicity of free nucleons, LCP's and IMF's as a function of scaled impact parameters for 
 $_{50}Sn^{124}+_{50}Sn^{124}$  and $_{50}Sn^{107}+_{50}Sn^{124}$ . 
It has been observed that as we move from central to peripheral collisions the number of free nucleons and LCP's decreases 
because the participation zone decreases which leads to the lower number of free nucleons and LCP's. But in case of IMF's, curve shows 
a rise and fall this is because for central collision the overlapping of participant and spectator zone is maximum so we get very small number 
of IMF's. For semi peripheral collisions the participant and spectator zone decreases so the production of IMF's increases and for peripheral collisions very small portion of target and projectile overlap so again few number of IMF's observed most of the fragments goes out as heavy mass fragments (HMF's). Moreover in Fig.1 we observe small production of free nucleons, LCP's and IMF's in $_{50}Sn^{107}+_{50}Sn^{124}$  as compare to $_{50}Sn^{124}+_{50}Sn^{124}$. This is because as the N/Z ratio increases the Coulomb repulsions increases which leads to the production of large number of products. The
equation\\
\begin{equation}
E(\rho)= E(\rho_{0})(\frac{\rho}{\rho_{0}})^\gamma
\end{equation}
 gives us the theoretical conjecture of how symmetry energy varies against density. $\gamma$, tells us the stiffness of the symmetry energy \cite{17}. In Fig.1 the free nucleons, LCP's and IMF's for $\gamma= 0$ and $\gamma= 0.66$ both the curves clearly indicates the density dependence of symmetry energy. Small difference is observed in both curves in case of LCP's at a scaled impact parameter range from 0.0 to 0.4. Moreover when we apply MSTP cut the number of free nucleons increases because at low impact parameter participant zone increases and large number of free nucleons produced on the other hand value of LCP's and IMF's decreases with MSTP cut. 

\begin{figure}
\includegraphics{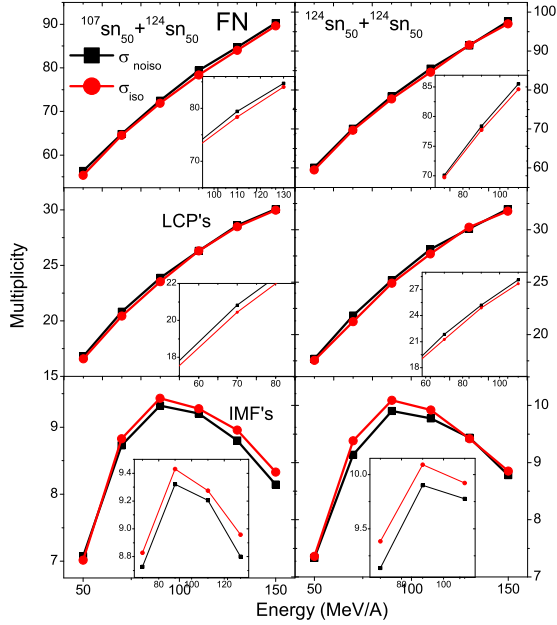}
\caption{\label{Fig:2} Multiplicity of free nucleons, LCP's and IMF's with energy at fixed scaled impact parameter for 
 $_{50}Sn^{124}+_{50}Sn^{124}$ and $_{50}Sn^{107}+_{50}Sn^{124}$.}
\end{figure}

Fig.\ref{Fig:2}, shows the variation of multiplicity of free nucleons, LCP's and IMF's with energy  at fixed scaled impact parameter for $_{50}Sn^{124}+_{50}Sn^{124}$ and $_{50}Sn^{107}+_{50}Sn^{124}$. It has been observed that multiplicity of free nucleons and LCP's increases with increase in energy. On the other hand, one can see a rise and fall in the multiplicity of IMF's; this behaviour is similar to the behaviour predicted by Aladin group \cite{18}. Moreover number of free nucleons and LCP's produced is very large as compare to IMF's this is because for central collision, collisions are violent so large number of free nucleons and LCP's produced as compare to IMF's. It is clear from the figure that slope of the curve is steeper  in case of $_{50}Sn^{124}+_{50}Sn^{124}$  than $_{50}Sn^{107}+_{50}Sn^{124}$ and this theoretical observation is in agreement with the experimental observation of Sfienti et al.\cite{18}. This rise is due to the fact that in case of neutron rich system, heavy residues with low excitation energy will predominantly emit neutrons, a channel that is suppressed in case of neutron-poor nuclei. We plot the curves for two different versions of the in-medium nucleon-nucleon cross-section term. The first one considers the different experimental neutron-proton, neutron-neutron, and proton-proton cross- section and is called the isospin dependent cross section $\sigma_{iso}$ \cite{15}. The second one, $\sigma_{noiso}$ , considers the same cross section for all the isospin channels \cite{13,14}. For these two cross sections we observe 0.80\% difference in free nucleons, 1.419\% in LCP's and 1.176\% in  case of IMF's for the reaction  $_{50}Sn^{107}+_{50}Sn^{124}$  and 0.86\% difference in free nucleons, 1.182\% in LCP's and 1.88\% in  case of IMF's for the reaction $_{50}Sn^{124}+_{50}Sn^{124}$. The difference is very small because the N/Z ratio is not far away from unity. But as N/Z increases the effect of different cross-section is clearly visible. The study on this topic is going on \cite{19}.

\begin{figure}
\includegraphics{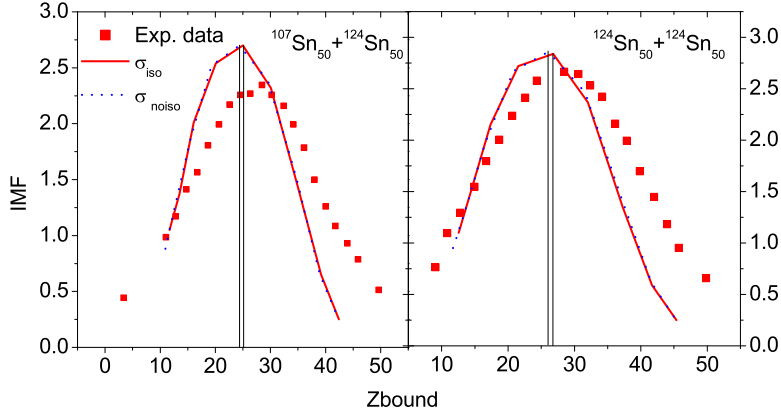}
\caption{\label{Fig:3} Multiplicity of IMF's as a function of $Z_{bound}$.}
\end{figure}

In Fig.\ref{Fig:3}, 
we have shown IMF's as a function of $Z_{bound}$. The quantity $Z_{bound}$ is defined as sum of all atomic charges $Z_{i}$ 
of all fragments with $Z_{i}> 2$. Here we observe that at semi peripheral collisions multiplicity IMF shows a peak because most of the spectator source does not take part in collision and large number of IMF's are observed. In case of central collision the collisions are violent so there few number of IMF's observed and for peripheral collisions very small portion of target and projectile overlap so again few number of IMF's observed most of the fragments goes out in heavy mass fragments (HMF's). In this way we get a clear "rise and fall" in multifragmentation emission. It is observed that IMF's shows the agreement with data at low impact parameters but fails at intermediate impact parameters due to no acess to filters. Moreover it has been observed that for the isospin dependent cross-section the curve shifts towards the experimental data for both  $_{50}Sn^{124}+_{50}Sn^{124}$ and 
 $_{50}Sn^{107}+_{50}Sn^{124}$ .\\

\begin{acknowledgments}
This work has been supported by a grant from the university grant commission, Government of India [Grant No. 39-858/2010 (SR)].\\
Authors are greatful to Dr. R. K. Puri for very fruitful discussion.\\

\end{acknowledgments}


\begin{thebibliography}{999}
\bibitem{1} A. D. Sood and R. K. Puri, Phys. Rev. C {\bf 69}, 054612(2004); A. D. Sood {\it et al.}, Phys. Lett. B {\bf 594}, 260(2004); A. D. Sood {\it et al.}, Eur. Phys. J. A {\bf 30}, 571(2006); A. D. Sood {\it et al.}, Phys. Rev. C {\bf 73}, 067602(2006); A. D. Sood {\it et al.}, Phys. Rev. C {\bf 79}, 064618(2009); R. Chugh and R. K. Puri, Phys. Rev. C {\bf 82}, 014603(2010). S.Kumar, M. K. Sharma, R. K. Puri, Phys. Rev. C {\bf 58}, 3494(1998); A. D. Sood and R. K .Puri, Phys. Rev. C {\bf 70}, 034611(2004); S. Gautam, A. D. Sood, R. K. Puri, J. Aichelin, Phys. Rev. C {\bf 83}, 014603(2011); ibid {\bf 83}, 034606(2011); S. Goyal and R. K. Puri Nucl. Phys. A {\bf 853}, 164(2011).
\bibitem{2} I. Dutt and R. K. Puri, Phys. Rev. C {\bf 81}, 047601(2010); ibid {\bf 81}, 044615(2010); ibid {\bf 81}, 064608(2010). 
\bibitem{3} Bao-An Li, Phys. Rev. C {\bf 67}, 017601(2003).
\bibitem{4} V. Greco {\it et al.,} Phys. Lett. B {\bf 562}, 215(2003).
\bibitem{5} W. P. Tan {\it et al.,} Phys. Rev. C {\bf 64}, 051901 (R)(2001).
\bibitem{6} M. B. Tsang {\it et al.,} Phys. Rev. Lett. {\bf 86}, 5023(2001).
\bibitem{7} H. Xu {\it et al.,} Phys. Rev. Lett.{\bf 85}, 716(2000).
\bibitem{8} J. F. Dempsey {\it et al.,} Phys. Rev. C {\bf 54}, 4(1996).
\bibitem{9} N. Buyukcizmeci {\it et al.,} Eur. Phys. Journal A {\bf 25}, 57(2005).
\bibitem{10} K. Schmidt {\it et al.,} Acta Physica Polonica B, {\bf 41}(2010).
\bibitem{11} M. B. Tsang {\it et al.,} Phys. Rev. Lett.{\bf 92}, 4(2004).
\bibitem{12} S. Gautam {\it et al.}, J. Phys. G {\bf 37}, 085102(2010).;
S. Kumar, S. Kumar and R. K. Puri, Phys. Rev. C {\bf 81}, 014601(2010);
S. Kumar, S. Kumar and R. K. Puri, Phys. Rev. C {\bf 81}, 014611(2010).
\bibitem{13} J. Aichelin, Phys. Rep. {\bf 202}, 233(1991); R. K. Puri, et al., Nucl. Phys. A {\bf 575}, 733(1994); ibid. J. Comp. Phys. {\bf 162}, 245(2000); E. Lehmann, R. K. Puri, A. Faessler, G. Batko, and 
S. W. Huang, Phys. Rev. C {\bf 51}, 2113(1995); ibid. Prog. Part. Nucl. Phys. {\bf 30}, 219(1993). Y. K. Vermani {\it et al.,} J. Phys. G. Nucl. Part. Phys. {\bf 36}, 0105103(2009); ibid {\bf 37}, 015105(2010); ibid Phys. Rev. C {\bf 79}, 064613(2009); ibid Nucl. Phys. A {\bf 847}, 243(2010).
\bibitem{14} S. Kumar, S. Kumar and R. K. Puri, Phys. Rev. C {\bf 78}, 064602(2008); ibid {\bf 57}, 2744(1998); S. Goyal and R. K. Puri, Phys. Rev. C {\bf 83}, 47601(2011).
\bibitem{15} C. Hartnack et al., Eur. Phys. J. A {\bf 1}, 151(1998); S.Kumar, S. Kumar and R. K. Puri, Phys. Rev. C {\bf 81}, 014611(2010); V. Kaur, S.Kumar and R. K. Puri, Phys. Lett. B, {\bf 697}, 512(2011); V. Kaur, S. Kumar and R. K. Puri, Nucl. Phys. A {\bf 861 }, 37(2011). 
\bibitem{16} S. Kumar and R. K. Puri, Phys. Rev.C {\bf 58}, 2858(1998); ibid {\bf 58}, 320(1998).
\bibitem{17} H. St\"ocker and W. Greiner, Phys. Rep. {\bf 137}, 277(1986); 
A. D. Sood and R. K. Puri, Phys.Rev.C {\bf 69}, 054612(2004).
\bibitem{18} C. Sfinti {\it et. al.,} Acta Physica Polonica B, {\bf 37}, 193(2006); ibid  Phys. Rev. Lett. 
{\bf 102}, 152701(2006).
\bibitem{19} A. Jain, S. Kumar and R. K. Puri,  Phys. Rev.C (To be submitted).
\end{thebibliography}
\end{document}